\documentstyle[12pt]{article}
\def\ie{{\em i.e.}}
\def\eg{{\em e.g.}}

\catcode`\@=11 

\def\lsim{\mathrel{\mathpalette\@versim<}}
\def\gsim{\mathrel{\mathpalette\@versim>}}
\def\@versim#1#2{\vcenter{\offinterlineskip
    \ialign{$\m@th#1\hfil##\hfil$\crcr#2\crcr\sim\crcr } }}

\def\t1{{\tilde 1}}

\def\h{\textstyle{1\over2}}

\def\to{\rightarrow}

\def\NPB#1#2#3{Nucl. Phys. B {\bf#1} (19#2) #3}
\def\PLB#1#2#3{Phys. Lett. B {\bf#1} (19#2) #3}

\def\PRT#1#2#3{Phys. Rep. {\bf#1} (19#2) #3}

\begin{document}

\begin{flushright}
CERN-TH.7259/94\\
CTP-TAMU-14/94\\
ACT-06/94\\
hep-th/9405120\\
\end{flushright}
\vglue 0.5cm
\begin{center}
{\Large\bf Moduli and K\"ahler potential in fermionic strings \\}
\vglue 1cm
JORGE L. LOPEZ$^{(a),(b)}$, D.V. NANOPOULOS$^{(a),(b),(c)}$,
and KAJIA YUAN$^{(a),(b)}$
\vglue 0.4cm
{\em $^{(a)}$Center for Theoretical Physics, Department of Physics\\
Texas A\&M University, College Station, TX 77843-4242, USA\\}
{\em $^{(b)}$Astroparticle Physics Group, Houston Advanced Research Center
(HARC)\\ The Woodlands, TX 77381, USA\\}
{\em $^{(c)}$CERN Theory Division, 1211 Geneva 23, Switzerland\\}
\vglue 1cm
{ABSTRACT}
\end{center}

\noindent We study the problem of identifying the moduli fields in fermionic
four-dimensional string models. We deform a free-fermionic model by introducing
exactly marginal operators in the form of Abelian Thirring interactions on the
world-sheet, and show that their couplings correspond to the untwisted moduli
fields. We study the consequences of this method for simple free-fermionic
models which correspond to $Z_2\times Z_2$ orbifolds and obtain their moduli
space and K\"ahler potential by symmetry arguments and by direct calculation
of string scattering amplitudes. We then generalize our analysis to more
complicated fermionic structures which arise in constructions of realistic
models corresponding to asymmetric orbifolds, and obtain the moduli space and
K\"ahler potential for this case. Finally we extend our analysis to the
untwisted matter sector and derive expressions for the full K\"ahler potential
to be used in phenomenological applications, and the target space duality
transformations of the corresponding untwisted matter fields.

\begin{flushleft}
CERN-TH.7259/94\\
CTP-TAMU-14/94\\
ACT-06/94\\
May 1994
\end{flushleft}
\vfill\eject
\setcounter{page}{1}
\pagestyle{plain}
\normalsize
\baselineskip=14pt

\section{Introduction}
Superstring theory remains the only consistent theoretical framework which
brings the expectation of unification of all fundamental interactions
including gravity. The study of the low-energy effective field theories
arising from four-dimensional string models, in particular those with $N=1$
space-time supersymmetry \cite{S4D}, is of crucial importance in bridging the
gap between string theory and the observed particle phenomenology, which at
the moment is embodied in the Standard Model. It is generally believed that
for energies well below the Planck scale, such string derived effective
theory should take the form of an $N=1$ locally supersymmetric quantum field
theory, namely, $N=1$ supergravity coupled to some $N=1$ Yang-Mills
supermultiplets and chiral supermultiplets. Therefore, its Lagrangian can be
specified in terms of three standard functions \cite{CFGP}: (i) the gauge
kinetic function $f_{ab}$, (ii) the superpotential $W$, and (iii) the
K\"ahler potential $K$. The ultimate goal would be to derive these functions
entirely from the underlying string theory. Indeed, much progress towards
this goal has been achieved in the past and there exists a great deal of
information about the structure of the effective Lagrangian for various
four-dimensional string models.

One of the advantages of string-derived effective field theories over
conventional $N=1$ supergravity theories, is the calculability of these three
functions ($f_{ab},W,K$) in string perturbation theory. This was demonstrated
in the $S$-matrix approach in Refs.~\cite{Smx1,Smx2,DKL}, where various
string scattering amplitudes were computed using the techniques of conformal
field theory. In addition, in the low-energy effective field theories of
string, a special class of massless fields, called moduli, play a unique
role. It is a distinct feature of string theory that some moduli fields
always exist. In $N=1$ supersymmetric string models, a modulus field is a
special massless chiral superfield which has flat scalar potential to all
orders in string perturbation theory, \ie, the vacuum expectation value (VEV)
of its scalar component is completely unconstrained. The VEVs of the moduli
parameterize a continuously connected family of string models. The study
of moduli fields and their symmetry properties, most notably target space
modular invariance or duality, has been found to be of great theoretical and
phenomenological interest, with applications in string-derived supergravities,
supersymmetry breaking, string threshold corrections to gauge couplings,
string cosmology, etc \cite{GPR}.

In this paper, we consider the low-energy effective field theory for a class
of four-dimensional heterotic string models constructed in the fermionic
formulation \cite{KLT,ABK}. The first studies of the effective field theory
for some simple models of this kind were performed sometime ago in
Refs.~\cite{AEFNT,FGKP}, by using a consistent truncation procedure similar
to the dimensional reduction of the ten-dimensional supergravity
Lagrangian \cite{DR10}. This procedure is based on the following two
observations: First, all $N=1$ string models in the fermionic formulation
can be obtained from an $N=4$ model with a gauge group of rank 22, by adding
non-trivial spin-structure vectors to reduce the space-time supersymmetry.
In the bosonic language, this corresponds precisely to constructing orbifold
models by introducing twists into  the generalized toroidal compactification
\cite{TCOM}. Second, the scalar couplings of $N=4$ supergravity possess
a unique $[SO(6,m)/SO(6)\times SO(m)]\otimes [SU(1,1)/U(1)]$ non-linear
sigma model structure \cite{N4SG}, the second factor corresponds to the
dilaton field. It was found in Refs.~\cite{AEFNT,FGKP} that, in addition to
the dilaton field which parameterizes $SU(1,1)/U(1)$, the {\it other}
scalar fields from the untwisted sector (the Neveu-Schwarz sector)
typically parameterize a K\"ahler manifold which is a direct product of
several factors of the form  $SO(2,n)/SO(2)\times SO(n)$.

This paper is organized as follows.
In Sec.~\ref{id} we establish a procedure with which the untwisted moduli
in fermionic models can be identified. We deform a free-fermionic model
by introducing exactly marginal operators in the form of world-sheet
Abelian Thirring interactions and show that their couplings correspond
to the untwisted moduli fields. In Sec.~\ref{simple} we illustrate the
procedure presented in Sec.~\ref{id} with simple fermionic models.
We identify explicitly the untwisted moduli fields, and obtain the
corresponding moduli space and K\"ahler potential by symmetry arguments.
The validity of the precise form of the K\"ahler potential is also
demonstrated with explicit string perturbation theory calculations.
In Sec.~\ref{z2} we expand on the connection between the simple
fermionic models in Sec.~\ref{simple} and symmetric
$Z_2\times Z_2$ orbifolds, focusing on the issue of untwisted moduli
fields in the two approaches. In Sec.~\ref{realistic} we generalize
our analysis to more complicated fermionic models corresponding to
asymmetric orbifolds which often arise in constructions of realistic
models. We show how the moduli space and K\"ahler potential can be
changed for this case with a typical example.
In Sec.~\ref{matter} we extend our discussion to the untwisted matter
fields and obtain the explicit form of the full K\"ahler potential for
all the untwisted scalar fields, which is particularly useful in
phenomenological studies of such fermionic string models.
In Sec.~\ref{target} we study the target space duality invariance
of the fermionic models, and as a by-product obtain the properties
of the untwisted matter fields under duality transformations.
Finally, we summarize our conclusions in Sec.~\ref{conclusions}.

\section{Identification of the untwisted moduli}
\label{id}
Given the importance of the moduli fields, first we would like to identify
the untwisted moduli in fermionic models. One very special untwisted
modulus field in such models, as in any string models, is the dilaton
field. Since the properties of the dilaton field are well-known, in this
paper we only discuss the other untwisted moduli fields in the
four-dimensional fermionic heterotic string models.
These models are described by two-dimensional {\em internal} conformal
and superconformal field theories of central charges $c_R=22$ and $c_L=9$
respectively, which in the fermionic formulation are completely fermionized.
It is particularly convenient to start with a model which makes use only of
{\it free} world-sheet fermions, since such a free-fermionic model can be
readily worked out using simple rules \cite{KLT,ABK} that ensure conformal
as well as modular invariance. Once such a free-fermionic model is
constructed, one would have found a special vacuum of the string theory,
which should correspond to a special point in the moduli space. Suppose
that one knows precisely what are the moduli fields in such free-fermionic
models, then by changing their vacuum expectation values one can obtain
other (continuously connected)
string vacua which may or may not be physically equivalent to the
original vacuum specified by the free-fermionic construction. However,
because the free-fermionic models have fixed (\ie, vanishing) values for
the moduli scalars, as opposed to the orbifold construction \cite{Orbi}
where arbitrary values of some moduli (such as the size of the orbifold)
are explicit, it is not evident which massless scalars correspond to
moduli fields in free-fermionic models.

For heterotic string models with $(2,2)$ internal superconformal theory,
there is a standard procedure \cite{Dixon,Sei,DKL} by which the moduli
fields can be unambiguously identified. This procedure can be applied to
the $(2,2)$ models constructed in the free-fermionic formulation, and entails
the existence of some massless scalar fields which can be viewed as the
scalar components of the moduli fields. However, since we are interested
in $(2,0)$\footnote{We use the convention in which the left-moving sector is
supersymmetric, whereas the right-moving sector is not.} string models that
can be constructed in the free-fermionic formulation, the above procedure
is not applicable. In fact, there seems to be no universal way of identifying
moduli in generic $(2,0)$ models \cite{DDxion}. Nevertheless, since
free-fermionic models can be viewed as orbifold models with special values
of the radii \cite{KLT}, and the untwisted moduli of various orbifold models
have been worked out and are valid for both $(2,2)$ and $(2,0)$ cases, we
are led to seek an analogy with the orbifold analysis.

It is crucial to recall that \cite{DVV}, in the context of conformal field
theory (CFT), moduli fields correspond to exactly marginal operators which
generate deformations of a CFT that preserve conformal invariance at the
classical as well as quantum level. For symmetric orbifold models, the
exactly marginal operators associated with the untwisted moduli fields
take the general form $\partial X^I{\bar \partial} X^J$, where
$X^I$ for $I=1,\ldots,6$ are the coordinates of the six-torus $T^6$.
Therefore, the untwisted moduli fields in such models admit the geometrical
interpretation of background fields \cite{NSW}, which appear as couplings
of the above exactly marginal operators in the non-linear sigma model
action that is the generating functional for string scattering
amplitudes \cite{CLO,CFG}. Based on this interpretation, the untwisted
moduli scalars are simply given by the background fields whose existence
preserves the point group symmetry of the corresponding orbifolds, and
the K\"ahler potential of these moduli fields or the metric of the
moduli space can be determined accordingly using symmetry
arguments \cite{CLO,CFG}.

Note that in the Frenkel-Kac-Segal construction \cite{FKS} of the Kac-Moody
current algebra from chiral bosons, the operator $i\partial X^I$ is nothing
but a $U(1)$ Cartan-subalgebra current. From the orbifold analogy, it is
therefore natural for us to expect that the exactly marginal operators in
the fermionic models should be given by {\it Abelian} Thirring operators
of the form $J^i_L(z){\bar J}^j_R({\bar z})$, where
$J^i_L, {\bar J}^j_R$ are some $U(1)$ chiral currents described by
world-sheet fermions. Indeed, it has been shown \cite{BNSY,KLT} that
Abelian Thirring interactions preserve conformal invariance, and a bosonic
string model with general background fields \cite{NSW} can be equivalently
obtained via the fermionic formulation by introducing general Thirring
interactions among the world-sheet fermions. In addition, a fermionic model
with non-vanishing Thirring interactions may be reformulated as a model
with only free world-sheet fermions with highly non-trivial
spin-structures \cite{BNSY}, but in this case the requirement of
modular invariance can not be simply solved as in Refs.~\cite{KLT,ABK}.
Thirring interactions have also been considered in the context of gauge
symmetry-breaking at low energies in fermionic models \cite{Hig,CHANG}.

In fact, even without referring to the orbifold analogy, it is not hard to
see that the Abelian Thirring operators $J^i_L(z){\bar J}^j_R({\bar z})$
satisfy the necessary and sufficient condition for integrability established
in Ref.~\cite{CHAU}, and hence they are exactly marginal. It is interesting
to note that the result of Ref.~\cite{CHAU} also implies the existence of
additional exactly marginal operators in certain fermionic models, and one
should regard the Abelian Thirring operators as a {\em minimal} set of
the exactly marginal operators generally available in fermionic models.
In this paper, we confine ourselves to the untwisted moduli associated
with this minimal set of exactly marginal operators, and we will use the
orbifold analogy throughout to illuminate our results.

Our focus here is to show how can one use the Abelian Thirring interactions
to identify the untwisted moduli in a class of fermionic models.
Contrary to the orbifold case where the modular invariant solutions
in the presence of non-trivial background fields and Wilson lines
are known \cite{IMNQ}, we do not attempt to find the most general
modular invariant solution with non-vanishing Thirring interactions
in the fermionic language. Instead, we follow the strategy of
Refs.~\cite{Hig,CHANG}. We start with a free-fermionic model with vanishing
Thirring interactions, and then perturb around this particular vacuum by
turning on some Thirring interactions. What we need in order to identify
the untwisted moduli fields are the symmetry properties of the Thirring
interactions. These can be determined in our perturbative analysis and are
expected to hold even for vacua far away from the original one.
{\em Our main observation} is that the untwisted moduli scalars in
fermionic models correspond to the Abelian Thirring interactions which,
if turned on, are compatible with the spin-structure of the free world-sheet
fermions. In essence, this is similar to the starting point of
Ref.~\cite{CLO}, \ie, that a background action becomes the generating
functional of string scattering amplitudes of a particular symmetric
orbifold model, if and only if it is invariant under the action of the
corresponding orbifold point group. We note that the analysis
of Ref.~\cite{CLO} cannot be directly used for the case of {\it asymmetric}
orbifolds \cite{asOrbi}, whereas our approach using Thirring interactions
seems to offer a viable method. In fact, in this paper we use this approach
to discuss the untwisted moduli in certain realistic fermionic models which
can only be interpreted as asymmetric orbifolds.

\section{Simple models: moduli and K\"ahler potential}
\label{simple}
To be concrete, in what follows we restrict ourselves to the fermionic
models whose spin-structure-vector bases ${\cal B}$ all contain a sub-basis
${\cal B}_{N=4}$, which by itself would generate an $N=4$ model with an
$SO(44)$ gauge group arising from the right-movers and an $U(1)^6$ gauge
group from the left-movers. For the sake of simplicity, we consider models
in which one can use six sets of left-moving real fermions
$(\chi^I,y^I,\omega^I)$ ($I=1,\ldots, 6$), each transforming in the adjoint
representation of $SU(2)$, to describe the left-moving $N=2$ superconformal
symmetry dictated by $N=1$ space-time supersymmetry.
For instance, we can choose ${\cal B}_{N=4}=\{{\bf 1},S\}$, where in
vector ${\bf 1}$ all world-sheet fermions are periodic, whereas in
the ``supersymmetry generating'' vector $S$ only the two transverse
$\psi^\mu$ and six $\chi^I$ world-sheet fermions are periodic and the
rest are antiperiodic. It is helpful to view these four-dimensional fermionic
models as the result of certain compactifications of the ten-dimensional
heterotic string. In this view, with the choice of $S$ just given, we can
identify the six $\chi^I$ with the fermionic superpartners of the six
compactified bosonic coordinates $X^I$ on the six-torus $T^6$. Therefore
each pair $(y^I,\omega^I)$ is nothing but the fermionized version of the
left-moving mode of the corresponding $X^I$ itself, \ie,
$i\partial X^I_L\sim y^Iw^I$. Clearly, the sector produced by the sub-basis
${\cal B}_{N=4}=\{{\bf 1},S\}$ corresponds to Narain's generalized toroidal
compactification \cite{TCOM}, and the sub-sector which is invariant under
orbifold twists gives rise to the untwisted sector of orbifold models.

Let us now consider the following two-dimensional action for the Abelian
Thirring interactions
\begin{equation}
S=\int d^2 z h_{ij}(X)J^i_L(z){\bar J}^j_R({\bar z}),
\label{eq:action}
\end{equation}
where $J^i_L$ ($i=1,\ldots,6$) are the chiral currents of the left-moving
$U(1)^6$, and $\bar J^j_R$ ($j=1,\ldots,22$) are the chiral currents of the
right-moving $U(1)^{22}$. The couplings $h_{ij}(X)$, as functions of the
space-time coordinates $X^\mu$, are four-dimensional scalar fields which we
will identify with the scalar components of the untwisted moduli
fields. In fact, for the simplest model with only the $N=4$ sub-basis
${\cal B}_{N=4}$, the $6\times 22$ fields $h_{ij}(X)$ in
Eq.~(\ref{eq:action}) are in one-to-one correspondence with the background
field metric $G_{IJ}$, the antisymmetric tensor $B_{IJ}$ ($I,J=1,\ldots,6$),
and the Wilson lines $A_{Ia}$ ($a=1,\ldots,16$).
These $h_{ij}(X)$ fields are precisely the moduli scalars of toroidal
compactification, which parameterize the coset space
$SO(6,22)/SO(6)\times SO(22)$ \cite{NSW}. This simple case was first
discussed in Ref.~\cite{KLT}.

We now point out that all these moduli scalars are indeed present
in the massless spectrum of the {\it free}-fermionic model given by
the basis $\{{\bf 1},S\}$. The massless scalar states of this model
are those from the Neveu-Schwarz sector, which transform in the adjoint
representation of $SO(44)$ and can be given in terms of 22
right-moving complex fermions ${\bar \Psi}^{+A}$ and their complex
conjugates ${\bar \Psi}^{-A}$ as ($1\leq A\not= B\leq 22$)
\begin{eqnarray}
|\chi^I\rangle &\otimes &|{\bar \Psi}^{+A}{\bar \Psi}^{-A}\rangle
\label{eq:ssa};\\
|\chi^I\rangle &\otimes &|{\bar \Psi}^{\pm A}{\bar \Psi}^{\pm B}\rangle
\label{eq:ssb}.
\end{eqnarray}
The $6\times22$ states in the Cartan subalgebra (those in (\ref{eq:ssa}))
are the massless moduli fields $h_{ij}(X)$ of this model
($i=I$ and $j=A$), and the corresponding marginal operators can be
given as (up to normalization constants)
\begin{equation}
J^i_L(z){\bar J}^j_R({\bar z})=:y^i(z)\omega^i(z):\
:{\bar \Psi}^{+j}({\bar z}){\bar \Psi}^{-j}({\bar z}):\ .
\label{eq:t1}
\end{equation}
In Eq.~(\ref{eq:t1}) the form of the left-moving chiral current $J^i_L(z)$ is
obtained from the fermionization of the six compactified coordinates
$i\partial X^I_L \sim y^I\omega^I$, which is also dictated by the
local $N=1$ internal world-sheet supercurrent
\begin{equation}
T^{\rm int}_F=i\sum_I\chi^Iy^I\omega^I.
\label{eq:superct}
\end{equation}
In fact, from the following OPE
\begin{equation}
T_F(w)\chi^I(z) \sim {1\over {w-z}}y^I(z)\omega^I(z),
\label{eq:ope}
\end{equation}
one can see that the marginal operator $J^i_L(z){\bar J}^j_R({\bar z})$
in Eq.~(\ref{eq:t1}) is the same as the zero-momentum vertex operator for
the associated moduli scalars $h_{ij}$ of Eq.~(\ref{eq:ssa}) in
the $0$-ghost picture.

We next consider a model with basis
${\cal B}=\{{\bf 1},S,b_1,b_2,b_3\}$, where
\footnote{See Ref.~\cite{Search} for our notation.}
\begin{eqnarray}
S&=(1\ 100\ 100\ 100\ 100\ 100\ 100\ :\
\overbrace{000000}^{\textstyle\bar y^I}\
\overbrace{000000}^{\textstyle\bar \omega^I}\
\!\!\overbrace{00000\ 000\ 0_8}^{\textstyle\bar\Psi^{\pm a}}),\\
b_1&=(1\ 100\ 100\ 010\ 010\ 010\ 010\ :\ 001111\ 000000\ 11111\ 100\
0_8),\\
b_2&=(1\ 010\ 010\ 100\ 100\ 001\ 001\ :\ 110000\ 000011\ 11111\ 010\
0_8),\\
b_3&=(1\ 001\ 001\ 001\ 001\ 100\ 100\ :\ 000000\ 111100\ 11111\ 001\
0_8).
\end{eqnarray}
In writing these spin-structure vectors we have separated the right-moving
fermions into 12 real fermions consisting of six pairs
${\bar y}^I, {\bar\omega}^I$ ($I=1,\ldots,6$), and the rest are treated as
16 complex fermions ${\bar \Psi}^{\pm a}$ ($a=1,\ldots,16$).
This separation amounts to a decomposition of the group $SO(44)$ into
its subgroup $SO(12)\times SO(32)$. With this separation we can now
use the bosonic analogy and consider the six pairs
${\bar y}^I, {\bar\omega}^I$ as the fermionization of the right-moving
modes of $X^I$. Therefore we can define
\begin{equation}
{\bar J}^j_R({\bar z})=:{\bar y}^j({\bar z}){\bar\omega}^j({\bar z}):
\quad (j=1,\ldots,6), \label{eq:rct1}
\end{equation}
and
\begin{equation}
{\bar J}^j_R({\bar z})=:{\bar \Psi}^{+(j-6)}({\bar z})
{\bar \Psi}^{-(j-6)}({\bar z}):\quad
(j=7,\ldots,22). \label{eq:rct2}
\end{equation}

We can also rewrite the massless states in the Cartan subalgebra (\ref{eq:ssa})
as two sets:
\begin{eqnarray}
(a)&\qquad |\chi^I\rangle\otimes|{\bar y}^I{\bar \omega}^I\rangle\qquad
&(I=1,\ldots,6) \label{eq:saa};\\
(b)&\qquad |\chi^I\rangle\otimes|{\bar \Psi}^{+a}{\bar \Psi}^{-a}\rangle
\quad &(a=1,\ldots,16) \label{eq:sbb},
\end{eqnarray}
which are the massless states in the Cartan subalgebras of $SO(12)$ and
$SO(32)$ respectively.
Clearly, using right-moving currents of the form in Eq.~(\ref{eq:rct1}) and
(\ref{eq:rct2}), we would get the same result for the simple $N=4$ model
as we did using (\ref{eq:t1}). In the models with additional spin-structure
basis vectors, some chiral currents ($J^i_L$ or ${\bar J}^j_R$) become
antiperiodic, and as a result certain terms in the general Thirring action
(\ref{eq:action}) are not invariant when the world-sheet fermions are
parallel-transported around the noncontractible loops of the world-sheet.
Such terms are inconsistent with the spin-structure of the fermions and are
therefore forbidden. We can view each additional spin-structure vector as
introducing a ``GSO-projection'' on the Thirring  action (\ref{eq:action}),
just as what these vectors do for physical string states, and only those terms
in action (\ref{eq:action}) which survive such ``GSO-projections'' lead to the
compatible exactly marginal operators and thus the untwisted moduli fields of
the model.

For example, from the form of $b_1$ it is easy to get the following
boundary conditions of the chiral currents:
\begin{eqnarray}
&J_L^{1,2} \rightarrow J_L^{1,2},\quad
&J_L^{3,4,5,6} \rightarrow -J_L^{3,4,5,6};\\
&{\bar J}_R^{1,2} \rightarrow {\bar J}_R^{1,2}, \quad
&{\bar J}_R^{3,4,5,6} \rightarrow -{\bar J}_R^{3,4,5,6},
\end{eqnarray}
whereas ${\bar J}_R^j (j=7,\ldots,22)$ are always periodic. Hence the
Thirring terms consistent with $b_1$ are simply
\begin{equation}
J_L^{1,2}{\bar J}_R^{1,2},\quad J_L^{3,4,5,6}{\bar J}_R^{3,4,5,6},\quad
 J_L^{1,2}{\bar J}_R^{j=7,\ldots,22}.
\label{eq:mo1}
\end{equation}
Following the symmetry argument of Ref.~\cite{CFG}, we see that for
model ${\cal B}=\{{\bf 1},S,b_1\}$, the untwisted
moduli scalars $h^{(1)}_{ij}$ $(i=1,2; j=1,2\ {\rm and}\ 7,\ldots,22)$
and $h^{(2)}_{ij}$ $(i,j=3,4,5,6)$ span a coset space
\begin{equation}
{\cal M}={SO(2,2+16)\over {SO(2)\times SO(2+16)}}\otimes
{SO(4,4)\over {SO(4)\times SO(4)}}.
\label{eq:space1}
\end{equation}
Working out the massless string states in Eqs.~(\ref{eq:saa}) and
(\ref{eq:sbb}) that survive the GSO-projection due to $b_1$, it is easy
to see that these states are indeed in one-to-one correspondence with the
marginal operators in Eq.~(\ref{eq:mo1}), as it should be. Again, these
marginal operators give the 0-ghost picture vertex operators of the
corresponding string states at zero momentum.

Carrying out the same analysis for $b_2$ and $b_3$, we find that the model
${\cal B}=\{{\bf 1},S,b_1,b_2,b_3\}$ has only the following allowed
Thirring terms
\begin{equation}
J_L^{1,2}{\bar J}_R^{1,2},\quad J_L^{3,4}{\bar J}_R^{3,4},\quad J_L^{5,6}
{\bar J}_R^{5,6},
\label{eq:mo2}
\end{equation}
and thus we get the following moduli space
\begin{equation}
{\cal M}={SO(2,2)\over {SO(2)\times SO(2)}}\otimes
{SO(2,2)\over {SO(2)\times SO(2)}}\otimes
{SO(2,2)\over {SO(2)\times SO(2)}},
\label{eq:space2}
\end{equation}
parameterized by the following three sets of untwisted moduli scalars
\begin{equation}
h_{ij}=|\chi^i\rangle\otimes|{\bar y}^j{\bar \omega}^j\rangle
\left\{
\begin{array}{ll}
(1)& (i,j=1,2)\\
(2)& (i,j=3,4)\\
(3)& (i,j=5,6)
\end{array}
\right.
\label{eq:stt}
\end{equation}

Thus far our results are consistent with those of Refs.~\cite{AEFNT,FGKP}.
However, in Refs.~\cite{AEFNT,FGKP}, apart from the dilaton field,
all the other scalar fields from the Neveu-Schwarz sector
were treated collectively, whereas we have been able to identify a
subset of these scalar fields (those given in Eq.~(\ref{eq:stt})) as the
untwisted moduli scalars, and the moduli space (\ref{eq:space2}) is just a
subspace of the full K\"ahler manifold underlying the non-linear sigma model of
the untwisted sector. In order to work out the low-energy effective field
theory for a string model, for instance the one given by ${\cal B}=\{{\bf
1},S,b_1,b_2,b_3\}$, it is necessary to know not only what K\"ahler manifold
the scalars span, but also {\it how} the scalars actually parameterize
the corresponding K\"ahler manifold. In other words, one would like to know
precisely the form of the K\"ahler potential or equivalently the K\"ahler
metric in terms of the scalar fields of concern. We next investigate this issue
for the untwisted moduli scalars given in Eq.~(\ref{eq:stt}).

Let us first briefly mention one property of the coset space
$SO(2,n)/SO(2)\times SO(n)$ ($n\geq 1$). One special parameterization of this
coset space, among many different possibilities \cite{Gilmore}, is to consider
it as a bounded subdomain of ${\bf C}^n$ that obeys the conditions
\footnote{Here we have chosen the complex coordinate system
$\alpha_i$ ($i=1,\ldots,n$) such that it is canonical at the origin.}
\begin{equation}
\left|\sum_i^n\alpha^2_i\right|<{1\over 2},\qquad
1-\sum_i^n\alpha_i{\bar\alpha}_i+{1\over 4}
\left|\sum_i^n\alpha^2_i\right|^2>0.
\end{equation}
In this parameterization, the coset space $SO(2,n)/SO(2)\times SO(n)$
assumes the following standard K\"ahler potential \cite{CV}
\begin{equation}
K(\alpha_i,{\bar\alpha}_i)=-\log\left(1-\sum_i^n\alpha_i{\bar\alpha}_i
+{1\over 4}\left|\sum_i^n\alpha^2_i\right|^2\right).
\label{eq:kkk}
\end{equation}
For what follows it is helpful to expand the K\"ahler potential in powers
of the fields. Up to fourth-order we get
\begin{equation}
K(\alpha_i,{\bar\alpha}_i)\approx \sum_i^n\alpha_i\bar\alpha_i
+{1\over4}\sum_i^n\alpha^2_i\bar\alpha^2_i
+\sum_{i<j}^n(\alpha_i\bar\alpha_i\alpha_j\bar\alpha_j
-{1\over4}\alpha^2_i\bar\alpha^2_j-{1\over4}\alpha^2_j\bar\alpha^2_i),
\label{eq:kkk4}
\end{equation}
and the second and fourth derivatives become
\footnote{We adopt the notation that subscripts on the K\"ahler potential
denote derivatives with respect to the corresponding fields, {\it e.g.},
$K_\Phi={\partial}_\Phi K$.}
\begin{equation}
K_{\alpha_i\bar\alpha_i}=1+\sum_j^n\alpha_j\bar\alpha_j,\qquad
K_{\alpha_i\bar\alpha_j}=-\alpha_i\bar\alpha_j+\bar\alpha_i\alpha_j,\ (i\not=j)
\end{equation}
and ($i\not=j$)
\begin{equation}
K_{\alpha_i\bar\alpha_i,\alpha_i\bar\alpha_i}=
K_{\alpha_i\bar\alpha_i,\alpha_j\bar\alpha_j}=
K_{\alpha_i\bar\alpha_j,\alpha_j\bar\alpha_i}=1,\qquad
K_{\alpha_i\bar\alpha_j,\alpha_i\bar\alpha_j}=-1\ .
\label{eq:kd}
\end{equation}

We now would like to show that the four real moduli scalars of each set
in Eq.~(\ref{eq:stt}) provide precisely the very special parameterization
of $SO(2,2)/SO(2)\times SO(2)$ in Eq.~(\ref{eq:kkk}). Because of the symmetric
structure of the moduli space (\ref{eq:space2}), we can just consider the first
set as an example. As it is often done in string calculations of this
class of models \cite{KLN}, we define naturally the following two
complex fields
\begin{eqnarray}
H^{(1)}_1=&{1\over\sqrt{2}}(h_{11}+ih_{21})
=&{1\over\sqrt{2}}|\chi^1+i\chi^2\rangle
\otimes|{\bar y}^1{\bar\omega}^1\rangle ,\label{eq:dd1}\\
H^{(1)}_2=&{1\over\sqrt{2}}(h_{12}+ih_{22})
=&{1\over\sqrt{2}}|\chi^1+i\chi^2\rangle
\otimes|{\bar y}^2{\bar\omega}^2\rangle .
\label{eq:dd2}
\end{eqnarray}
Then, using the methods of Ref.~\cite{KLN}, we obtain the following
non-vanishing string scattering amplitudes
\begin{eqnarray}
{\cal A}(H_1,H_2,{\bar H}_2,{\bar H}_1)&=&
{\cal A}(H_2,H_1,{\bar H}_1,{\bar H}_2)\nonumber\\
&=&-{g^2\over 4}\ {{\Gamma(-s/8)\Gamma(-t/8)\Gamma(-u/8)}\over
{\Gamma(s/8)\Gamma(t/8)\Gamma(u/8)}}
\left\{{s\over {1+t/8}}+{su\over {t(1+t/8)}}\right\},\nonumber\\
&&\\
\label{eq:ap1}
{\cal A}(H_1,H_2,{\bar H}_1,{\bar H}_2)&=&
{\cal A}(H_2,H_1,{\bar H}_2,{\bar H}_1)\nonumber\\
&=&-{g^2\over 4}\ {{\Gamma(-s/8)\Gamma(-t/8)\Gamma(-u/8)}\over
{\Gamma(s/8)\Gamma(t/8)\Gamma(u/8)}}
\left\{{s\over {1+u/8}}+{st\over {u(1+u/8)}}\right\},\nonumber\\
&&\\
\label{eq:ap2}
{\cal A}(H_1,H_1,{\bar H}_1,{\bar H}_1)&=&
{\cal A}(H_2,H_2,{\bar H}_2,{\bar H}_2)\nonumber\\
&=&-{g^2\over 4}\ {{\Gamma(-s/8)\Gamma(-t/8)\Gamma(-u/8)}\over
{\Gamma(s/8)\Gamma(t/8)\Gamma(u/8)}}\nonumber\\
&&\left\{{s\over {1+u/8}}+{s\over {1+t/8}}+{st\over {u(1+u/8)}}
+{su\over {t(1+t/8)}}-{s\over {1+s/8}}\right\},\nonumber\\
&&\\
\label{eq:ap3}
{\cal A}(H_1,H_1,{\bar H}_2,{\bar H}_2)&=&
{\cal A}(H_2,H_2,{\bar H}_1,{\bar H}_1)\nonumber\\
&=&{g^2\over 4}\ {{\Gamma(-s/8)\Gamma(-t/8)\Gamma(-u/8)}\over
{\Gamma(s/8)\Gamma(t/8)\Gamma(u/8)}}\left\{{s\over {1+s/8}}\right\}.
\label{eq:ap4}
\end{eqnarray}
To quadratic order in the momenta, Eqs.~(\ref{eq:ap1})--(\ref{eq:ap4})
become
\begin{eqnarray}
{\cal A}(H_1,H_2,{\bar H}_2,{\bar H}_1)&=&
{\cal A}(H_2,H_1,{\bar H}_1,{\bar H}_2)=
{g^2\over 4}\left({su\over t}+s\right),
\label{eq:ap1p}\\
{\cal A}(H_1,H_2,{\bar H}_1,{\bar H}_2)&=&
{\cal A}(H_2,H_1,{\bar H}_2,{\bar H}_1)=
{g^2\over 4}\left({st\over u}+s\right),
\label{eq:ap2p}\\
{\cal A}(H_1,H_1,{\bar H}_1,{\bar H}_1)&=&
{\cal A}(H_2,H_2,{\bar H}_2,{\bar H}_2)=
{g^2\over 4}\left({st\over u}+{su\over t}+s\right),
\label{eq:ap3p}\\
{\cal A}(H_1,H_1,{\bar H}_2,{\bar H}_2)&=&
{\cal A}(H_2,H_2,{\bar H}_1,{\bar H}_1)=
{g^2\over 4}(-s).
\label{eq:ap4p}
\end{eqnarray}
We now compare the string scattering amplitudes
(\ref{eq:ap1p})--(\ref{eq:ap4p}) with those that would be obtained from $N=1$
supergravity calculations \cite{DKL,BGL}. For this purpose, we only need
the scattering amplitudes due to sigma-model interactions as well as
gravity, namely
\begin{equation}
{\cal A}(H_i,H_j,{\bar H}_k,{\bar H}_l)\propto\left(
{su\over t}\,\delta_{il}\delta_{jk}+{st\over u}\,
\delta_{ik}\delta_{jl}+s K_{H_j\bar H_k,H_i\bar H_l}\right).
\label{eq:asugra}
\end{equation}
This comparison yields
\begin{eqnarray}
K_{H_1{\bar H}_1,H_2{\bar H}_2}&=&
K_{H_2{\bar H}_2,H_1{\bar H}_1}=
K_{H_1{\bar H}_1,H_1{\bar H}_1}=
K_{H_2{\bar H}_2,H_2{\bar H}_2}=1\;,\label{eq:K0K1}\\
K_{H_1{\bar H}_2,H_2{\bar H}_1}&=&
K_{H_2{\bar H}_1,H_1{\bar H}_2}=1\;,\label{eq:K0K2}\\
K_{H_1{\bar H}_2,H_1{\bar H}_2}&=&
K_{H_2{\bar H}_1,H_2{\bar H}_1}=-1\;,\label{eq:K0K3}
\end{eqnarray}
which is consistent with Eq.~(\ref{eq:kd}) for the case of $n=2$,
and $\alpha_{1,2}=H_{1,2}$. We note in passing that, since in our
current case the exactly marginal operators corresponding
to the untwisted moduli fields are known, one can also
perturbatively compute the K\"ahler metric of the moduli
space by calculating the Zamolodchikov metric \cite{ZLD} with
these exactly marginal operators, which we expect will lead
to the same results.

{}From the above analysis,
for the model ${\cal B}=\{{\bf 1},S,b_1,b_2,b_3\}$,
the K\"ahler potential of the untwisted moduli
takes the following explicit form
\begin{equation}
K(H,{\bar H})=-\sum^3_{i=1}\log\left(1-\sum_{j=1,2}H^{(i)}_j{\bar H}^{(i)}_j
+{1\over 4}\left|\sum_{j=1,2}H^{(i)}_jH^{(i)}_j\right|^2\right),
\label{eq:kp}
\end{equation}
where the complex fields $H^{(2)}_{1,2}$ and $H^{(3)}_{1,2}$ are
defined in a similar fashion as Eqs.~(\ref{eq:dd1}) and (\ref{eq:dd2}).

\section{Comparison with $Z_2\times Z_2$ orbifolds}
\label{z2}
It is instructive to revisit the results in Sec.~\ref{simple} for the
fermionic string models in light of the orbifold analogy. First of all,
the coset space $SO(2,2)/SO(2)\times SO(2)$ is a reducible K\"ahler
manifold, since \cite{CV}
\begin{equation}
{SO(2,2)\over {SO(2)\times SO(2)}}\simeq
{SU(1,1)\over U(1)}\otimes {SU(1,1)\over U(1)}.
\label{eq:ism1}
\end{equation}
Therefore, the untwisted moduli space (\ref{eq:space2}) for the model
${\cal B}=\{{\bf 1},S,b_1,b_2,b_3\}$ can be written in the following
more familiar form
\begin{equation}
{\cal M}=\left[{SU(1,1)\over U(1)}\otimes {SU(1,1)\over U(1)}\right]^3.
\label{eq:stspc}
\end{equation}
This is exactly the untwisted moduli space for the symmetric $Z_2\times Z_2$
orbifold model \cite{IBLust}, which can be obtained by using the method of
Ref.~\cite{CLO}. Furthermore, we can make a linear transformation which maps
each pair of the original complex string fields $H^{(i)}_{1,2}$ into another
pair $\Phi^{(i)}_{T,U}$. For the first set we have
\begin{eqnarray}
\Phi^{(1)}_T=&{1\over\sqrt{2}}(H^{(1)}_1-iH^{(1)}_2)
=&{1\over\sqrt{2}}|\chi^1+i\chi^2\rangle
\otimes {1\over\sqrt{2}}
|{\bar y}^1{\bar\omega}^1-i{\bar y}^2{\bar\omega}^2\rangle ,\label{eq:T}\\
\Phi^{(1)}_U=&{1\over\sqrt{2}}(H^{(1)}_1+iH^{(1)}_2)
=&{1\over\sqrt{2}}|\chi^1+i\chi^2\rangle
\otimes {1\over\sqrt{2}}
|{\bar y}^1{\bar\omega}^1+i{\bar y}^2{\bar\omega}^2\rangle .\label{eq:U}
\end{eqnarray}
The other two sets, $\Phi^{(2)}_{T,U}$ and $\Phi^{(3)}_{T,U}$, are defined
analogously. Note that the $T$-type fields are {\it not} the complex
conjugates of the corresponding $U$-type fields; they are two independent
complex fields.

In terms of the new complex fields $\Phi^{(i)}_{T,U}$, the K\"ahler
potential of the untwisted moduli fields of the model
${\cal B}=\{{\bf 1},S,b_1,b_2,b_3\}$, given in Eq.~(\ref{eq:kp})
in Sec.~\ref{simple}, can be rewritten as
\begin{equation}
K(\Phi,{\bar\Phi})=-\sum^3_{i=1}
\log\left(1-\Phi^{(i)}_T{\bar\Phi}^{(i)}_T\right)
-\sum^3_{i=1}\log\left(1-\Phi^{(i)}_U{\bar\Phi}^{(i)}_U\right).
\label{eq:kp1}
\end{equation}
In Eq.~(\ref{eq:kp1}) the $T$-type and $U$-type fields are completely
separated, and each $T$-type (or $U$-type) field provides a bounded
parameterization of the coset space $SU(1,1)/U(1)$. Of course, the reason
that we can define complex moduli fields $\Phi^{(i)}_{T,U}$ with this
property for our fermionic model is precisely the isomorphism (\ref{eq:ism1}).

Once again, Eq.~(\ref{eq:kp1}) can be confirmed by computing the string
scattering amplitudes of four moduli fields, as we did in Sec.~\ref{simple}.
This time we find that the non-vanishing string scattering amplitude is
the one involving the same four $T$-type (or $U$-type) fields, given by
\begin{eqnarray}
{\cal A}(\Phi,\Phi,{\bar\Phi},{\bar\Phi})&=&
-{g^2\over 4}\ {{\Gamma(-s/8)\Gamma(-t/8)\Gamma(-u/8)}\over
{\Gamma(s/8)\Gamma(t/8)\Gamma(u/8)}}\nonumber\\
&&\times\left\{{s\over {1+u/8}}+{s\over {1+t/8}}+{st\over {u(1+u/8)}}
+{su\over {t(1+t/8)}}\right\},
\end{eqnarray}
which in the low-energy limit becomes (to quadratic order in the momenta)
\begin{equation}
{\cal A}(\Phi,\Phi,{\bar\Phi},{\bar\Phi})=
{g^2\over 4}\left({st\over u}+{su\over t}+2s\right).
\label{eq:4pt}
\end{equation}
{}From Eq.~(\ref{eq:4pt}) one can infer that the metric of the moduli space
has only the following non-vanishing component \cite{BGL}
\begin{equation}
K_{\Phi\bar\Phi}=1+2\Phi\bar\Phi\approx{1\over ({1-\Phi{\bar\Phi}})^2}\ ,
\label{eq:kahler}
\end{equation}
which is exactly the standard Fubini--Study metric of $SU(1,1)/U(1)$
derived from the K\"ahler potential (\ref{eq:kp1}).

The above results are expected because our fermionic model with
$b_1$, $b_2$ and $b_3$ can be regarded as a symmetric
$Z_2\times Z_2$ orbifold model. It was first noted in Ref.~\cite{KLT}
that in the type of fermionic models we are considering there is a
$Z_2$ orbifold structure. The connection between such fermionic models
and the $Z_2\times Z_2$ orbifold model was explained in
Ref.~\cite{IBtalk} (see also Ref.~\cite{Alon}). To clearly see this,
let us first carry out explicitly the bosonization described in
Sec.~\ref{simple},
\begin{equation}
e^{iX^I_L}={1\over \sqrt{2}}(y^I+i\omega^I),\quad
e^{iX^I_R}={1\over \sqrt{2}}({\bar y}^I+i{\bar \omega}^I)
\quad(I=1,\ldots,6).
\label{eq:boson}
\end{equation}
We then form the three complex planes as follows ($X=X_L+X_R$)
\begin{equation}
Z^{\pm}_k={1\over \sqrt{2}}(X^{2k-1} \pm iX^{2k}),\quad
\psi^{\pm}_k={1\over \sqrt{2}}(\chi^{2k-1} \pm i\chi^{2k})
\quad (k=1,2,3),
\label{eq:complex}
\end{equation}
where the $Z^\pm_k$ are the complex coordinates of the six compactified
dimension now viewed as three complex planes, and $\psi^{\pm}_k$ are the
corresponding superpartners. From (\ref{eq:boson}) and (\ref{eq:complex}),
one see that $b_1$ can be interpreted as the twist $\theta$ of a
symmetric $Z_2\times Z_2$ orbifold which keeps the first complex plane
unrotated, but rotates the second and the third ones simultaneously by $\pi$.
Indeed, under $b_1$ we have $y^{1,2}\to-y^{1,2}$,
$\omega^{1,2}\to-\omega^{1,2}$,
and thus $e^{iX^{1,2}_L}\to-e^{iX^{1,2}_L}=e^{i(X^{1,2}_L+\pi)}$
Whereas $y^{3,4,5,6}\to y^{3,4,5,6}$,
$\omega^{3,4,5,6}\to-\omega^{3,4,5,6}$,
and thus $e^{iX^{3,4,5,6}_L}\to e^{-iX^{3,4,5,6}_L}$.
The right-moving modes have the same transformations, because of
the form of $b_1$ chosen. This fact indicates that the orbifold
is symmetric. Therefore, the first complex plane is only shifted,
{\it i.e.}, $Z^\pm_1\to Z^\pm_1+{\rm shift}$, but the
other two are rotated, {\it i.e.},
$Z^\pm_{2,3}\to -Z^\pm_{2,3}=e^{i\pi}Z^\pm_{2,3}$.
Also, from $\chi^{1,2}\to\chi^{1,2}$ and
$\chi^{3,4,5,6}\to-\chi^{3,4,5,6}$ one
gets $\psi^\pm_1\to\psi^\pm_1$ and $\psi^\pm_{2,3}\to-\psi^\pm_{2,3}$.
Analogously, one can show that $b_2$ is the twist $\omega$ which rotates
the first and the third complex planes by $\pi$, and finally, $b_3$ is
the twist $\theta\omega$ that keeps the third complex plane fixed. As
mentioned in Sec.~\ref{simple}, it is this connection between fermionic
models and orbifold models that inspired our approach.

Let us make use of this connection a bit further. In terms of the complex
coordinates $Z^\pm_k$, we can rewrite the allowed Thirring interaction
terms for the untwisted moduli of Eq.~(\ref{eq:stt}) in bosonic form.
For instance, we have for the first set
\begin{equation}
\sum_{i,j=1,2}h_{ij}J^i_L{\bar J}^j_R
=\Phi^{(1)}_T\partial Z^{-}_1{\bar\partial} Z^{+}_1
+{\bar\Phi}^{(1)}_T\partial Z^{+}_1{\bar\partial} Z^{-}_1
+\Phi^{(1)}_U\partial Z^{-}_1{\bar\partial} Z^{-}_1
+{\bar\Phi}^{(1)}_U\partial Z^{+}_1{\bar\partial} Z^{+}_1\; ,
\label{eq:ccc}
\end{equation}
where $\Phi^{(1)}_{T,U}$ are the complex fields defined in
Eqs.~(\ref{eq:T}) and (\ref{eq:U}).
Obviously the same can be done for the other
two sets. From Eq.~(\ref{eq:ccc}), we immediately see that the $T$-type
moduli field $\Phi_T$ is associated with the exactly marginal operator
$\partial Z^{-}{\bar\partial} Z^{+}$ which deforms the K\"ahler class of
the compact space, and the $U$-type moduli field $\Phi_U$ is associated with
the operator $\partial Z^{-}{\bar\partial} Z^{-}$ which deforms the complex
structure of the compact space. Therefore, the above orbifold analogy allows
us to assign the geometrical meaning to the $T$-type moduli as those
corresponding to the harmonic $(1,1)$ forms of the compact space, and the
$U$-moduli as those corresponding to the $(2,1)$ forms. The number of such
fields are given by the non-trivial Hodge numbers of the compact space,
which for the case of the symmetric $Z_2\times Z_2$ orbifold are
$h^{(1,1)}=3$ and $h^{(2,1)}=3$ \cite{IBLust}. This is the
reason why we get  precisely three sets of $\Phi^{(i)}_T$ and three sets
of $\Phi^{(i)}_U$ in the model ${\cal B}=\{{\bf 1},S,b_1,b_2,b_3\}$.
In fact, from the previous discussion it should be clear that the
properties of the untwisted moduli which we have derived are not only
valid in this simple model, but remain valid in any fermionic model with
additional spin-structure vectors, {\it as long as} these new vectors
{\it do not} spoil the symmetric $Z_2\times Z_2$ orbifold structure.
Examples of such fermionic models can be found, {\it e.g.}, in Ref.~\cite{Joe}.

\section{A new feature in realistic models}
\label{realistic}
Phenomenologically realistic fermionic models have been constructed by
adding more spin-structure vectors to the basis ${\cal B}$
\cite{fsu5re,fsu5old,pati,smlike,Search}. In these models,
there are spin-structure vectors which assign {\it asymmetrically}
the boundary conditions for the left-moving real fermions $y^I,\omega^I$
relative to the right-moving ones ${\bar y}^I,{\bar \omega}^I$, so that the
left-moving mode of some compactified coordinate $X^I_L$ and the
corresponding right-moving mode $X^I_R$ will be twisted differently.
{}From the discussion in Sec.~\ref{z2}, it is clear that such
models should be interpreted as asymmetric orbifolds \cite{asOrbi}.
We now address this new feature of such models in connection with
the determination of the untwisted moduli fields.

It is convenient to consider a concrete and typical example, for which
we choose the ``revamped'' flipped $SU(5)$ model \cite{fsu5re}.
In this model, the ``asymmetry'' of the twist is introduced by the
following basis vector
\begin{equation}
\alpha=(0\ 000\ 000\ 000\ 011\ 000\ 011:\ 000101\ 011101\
\h\h\h\h\h\ \h\h\h\ \h\h\h\h 1100).
\end{equation}
Now consider the remaining Thirring terms in Eq.~(\ref{eq:mo2}). Under this
vector $\alpha$, the left-moving current $J^2_L$ is periodic but its
right-moving counterpart ${\bar J}^2_R$ is antiperiodic; the same holds
for $J^3_L$ and $\bar J^3_R$. Note also that currents $J^1_L$ and $J^4_L$
remain periodic. As a result, operators $J^{1,2}_L{\bar J}^2_R$
and $J^{3,4}_L{\bar J}^3_R$ and no longer consistent with the spin-structure
of the model that contains the vector $\alpha$. In fact, it is easy to check
that the string states $h_{12}$, $h_{22}$, $h_{33}$ and $h_{43}$ that were
present in Eq.~(\ref{eq:stt}) do not exist in the massless spectrum of this
model any more. They have been projected out of the spectrum by precisely
the GSO-projection due to $\alpha$.

In this model, therefore, the remaining Thirring terms are
\begin{equation}
J_L^{1,2}{\bar J}_R^{1},\quad J_L^{3,4}{\bar J}_R^{4},
\quad J_L^{5,6}{\bar J}_R^{5,6},
\label{eq:mo3}
\end{equation}
and thus the untwisted moduli space reduces to
\begin{equation}
{\cal M}={SO(2,1)\over SO(2)}\otimes
{SO(2,1)\over SO(2)}\otimes
{SO(2,2)\over {SO(2)\times SO(2)}},
\label{eq:space3}
\end{equation}
which because of the isomorphisms (\ref{eq:ism1}) and \cite{CV}
\begin{equation}
{SO(2,1)\over SO(2)}\simeq {SU(1,1)\over U(1)},
\label{eq:ism2}
\end{equation}
is isomorphic to
\begin{equation}
{\cal M}={SU(1,1)\over U(1)}\otimes {SU(1,1)\over U(1)}\otimes
\left[{SU(1,1)\over U(1)}\otimes {SU(1,1)\over U(1)}\right].
\label{eq:stf5}
\end{equation}
Using the complex notation $H^{(i)}_{1,2}$ defined in
Sec.~\ref{simple}, we see that the untwisted moduli fields
for this model are just
\begin{equation}
H^{(1)}_1,\qquad H^{(2)}_2, \qquad H^{(3)}_{1,2},
\end{equation}
and they give the following K\"ahler potential (see Eq.~(\ref{eq:kp}))
\begin{eqnarray}
K(H,{\bar H})&=&-2\log\left(1-{1\over 2}H^{(1)}_1{\bar H}^{(1)}_1\right)
-2\log\left(1-{1\over 2}H^{(2)}_2{\bar H}^{(2)}_2\right)\nonumber\\
&&-\log\left(1-\sum_{j=1,2}H^{(3)}_j{\bar H}^{(3)}_j
+{1\over 4}\left|\sum_{j=1,2}H^{(3)}_jH^{(3)}_j\right|^2\right).
\label{eq:fkp}
\end{eqnarray}

In this model, the third set in Eq.~(\ref{eq:stt}) is intact, which gives
one $T$-type field $\Phi^{(3)}_T$ and one $U$-type field $\Phi^{(3)}_U$.
But the first and second sets each give only one complex modulus field,
and one can no longer attribute to them the geometrical meaning of
either $T$-type ($(1,1)$ form) or $U$-type ($(2,1)$ form), simply because
now the argument which leads to Eq.~(\ref{eq:ccc}) does not apply. However,
because of the isomorphism (\ref{eq:ism2}), it should still be possible to
find a new parameterization of $SO(2,1)/SO(2)$, such that the first two terms
in Eq.~(\ref{eq:fkp}) can be recast into the Fubini-Study form of
$SU(1,1)/U(1)$. The desired parameterization is provided by the
following transformation
\begin{equation}
\Phi=H\left(1-{1\over 4}H{\bar H}\right)^{1\over 2}.
\label{eq:nonli}
\end{equation}
Transforming $H^{(1)}_1$ ($H^{(2)}_2$) into $\Phi_1$ ($\Phi_2$)
according to (\ref{eq:nonli}), and also denoting $\Phi^{(3)}_{T,U}$
simply by $\Phi_{3,4}$, the K\"ahler potential (\ref{eq:fkp}) becomes
\begin{equation}
K(\Phi,{\bar\Phi})=-\sum^4_{i=1}\log\left(1-\Phi_i{\bar\Phi}_i\right),
\label{eq:kp2}
\end{equation}
to be contrasted with that obtained before introducing the vector $\alpha$,
\ie, Eq.~(\ref{eq:kp1}).

\section{The untwisted matter fields}
\label{matter}
In addition to the untwisted moduli which correspond to the Cartan subalgebra
at the $N=4$ level (see Eq.(\ref{eq:ssa})), there are other massless scalar
fields in the Neveu-Schwarz sector which correspond to the non-zero roots of
$D_{22}$ at the $N=4$ level (see Eq.(\ref{eq:ssb})). Contrary to the untwisted
moduli, these scalar fields in general are not associated with exactly
marginal operators, and for this reason they should be treated as untwisted
matter fields. For the class of $N=1$ fermionic string models we are
considering, as we have shown in previous sections, the untwisted moduli fields
split up into three sets, each of which parameterizes a coset space:
$SO(2,2)/SO(2)\times SO(2)$ or $SO(2,1)/SO(2)$. Similarly, in such $N=1$ models
the untwisted matter fields that survive the various GSO-projections also fall
into three sets. It is interesting to investigate the non-linear sigma model
structure of these untwisted matter fields.

This problem in fact has already been partially solved. In
Refs.~\cite{AEFNT,FGKP}, by using a truncation method, it was shown that each
set of scalar fields from the Neveu-Schwarz sector admits a non-linear sigma
model structure of $SO(2,n)/SO(2)\times SO(n)$, where $n$ counts the total
number of the scalar fields in the set, which we now know includes {\it both}
moduli {\it and} matter fields. Furthermore, the K\"ahler potential of the form
(\ref{eq:kkk}) was also written down in Ref.~\cite{FGKP} by solving the
constraints satisfied by the representative fields for the $N=4$ matter scalar
manifold \cite{N4SG}, although the complex coordinate system used there is not
canonical at the origin. Ref.~\cite{AEFNT} obtained another form of the
K\"ahler potential by solving the constraints in a slightly different way. What
was {\em not} addressed in these papers is the issue of how the canonical
complex coordinates $\alpha_i$ ($i=1,\ldots,n$) which appear in the
K\"ahler potential (\ref{eq:kkk}) are related to the actual
massless string states. We now elucidate this relation.

In Sec.~\ref{simple}, we established this relation for the moduli fields by
calculating various string scattering amplitudes. In this case the complex
coordinates $\alpha_{1,2}$ are simply given by the string states $H_{1,2}$
defined in Eqs.~(\ref{eq:dd1}) and (\ref{eq:dd2}). For the matter fields one
would expect that such direct relations also hold. That is, one can just
write down a string state in the most natural way, and then identify it with a
coordinate $\alpha$. Interestingly enough, we found that this is not always the
case. For matter states whose right-moving part consists of only real
fermionic oscillators, {\it e.g.}, a state with
$|{\bar y}^I{\bar\omega}^J\rangle$ ($I\not= J$), the string states themselves
give the coordinates $\alpha$'s. However, it is quite common for matter states
to have the right-moving part consist of complex fermionic oscillators, of the
form $|{\bar\Psi}^{\pm a}{\bar\Psi}^{\pm b}\rangle$ ($a\not= b$),
$|{\bar y}^I{\bar\Psi}^{\pm a}\rangle$, or
$|{\bar\omega}^I{\bar\Psi}^{\pm a}\rangle$. Such matter fields always come in
pairs, such that the right-moving oscillators in each pair are complex
conjugates of each other. (The fields themselves are distinct since their
left-moving oscillators are the same.) We found that for each such pair of
matter fields, the two corresponding canonical complex coordinates entering in
the K\"ahler potential (\ref{eq:kkk}) are given by the {\it real} and {\it
imaginary} parts of the right-moving oscillators respectively, with the complex
left-moving part untouched. In this sense, one can regard such pairs of string
matter states as corresponding to the pair $\Phi_{T,U}$ in (\ref{eq:T}) and
(\ref{eq:U}), and then the procedure of finding the canonical complex
coordinates simply corresponds to finding $H_{1,2}$ in terms of $\Phi_{T,U}$
from Eqs.~(\ref{eq:T}) and (\ref{eq:U}).

We now illustrate the above discussion in the ``revamped''
flipped $SU(5)$ model \cite{fsu5re}. In the notation of Ref.~\cite{fsu5re},
the matter fields in the {\em first set} of this model are:
($a=1,\ldots,5$)
\begin{eqnarray}
\Phi_{23}=&{1\over\sqrt{2}}|\chi^1+i\chi^2\rangle\otimes
|{\bar\Psi}^{-7}{\bar\Psi}^{+8}\rangle,\quad
{\bar\Phi}_{23}=&{1\over\sqrt{2}}|\chi^1+i\chi^2\rangle\otimes
|{\bar\Psi}^{+7}{\bar\Psi}^{-8}\rangle,\label{eq:f23}\\
h_1=&{1\over\sqrt{2}}|\chi^1+i\chi^2\rangle\otimes
|{\bar\Psi}^{-a}{\bar\Psi}^{+6}\rangle,\quad
{\bar h}_1=&{1\over\sqrt{2}}|\chi^1+i\chi^2\rangle\otimes
|{\bar\Psi}^{+a}{\bar\Psi}^{-6}\rangle.\label{eq:hh}
\end{eqnarray}
Note that ${\bar h}_1$ is {\it not} the complex conjugate of $h_1$: under
$SU(5)\times U(1)$ $h_1$ transforms as $({\bf 5},1)$, whereas ${\bar h}_1$
transforms as $({\bar{\bf 5}},-1)$. Similarly, $\Phi_{23}$ and
${\bar\Phi}_{23}$ are different fields, which transform as singlets of
$SU(5)\times U(1)$, but carry additional $U(1)$ charges. In what follows,
when discussing specifically about this model,  we denote the complex conjugate
of field $\Phi$ by $\Phi^\dagger$ (instead of ${\bar\Phi}$) to avoid possible
confusions with the notation. The untwisted matter fields in the first set
consist of $12=5+5+1+1$ complex degrees of freedom, which combined with the
one modulus field $\Phi_1=H^{(1)}_1$ (see Sec.~\ref{realistic}), altogether
parameterize a coset space $SO(2,1+12)/[SO(2)\times SO(1+12)]$.

Let us consider the matter fields $\Phi_{23}$ and ${\bar\Phi}_{23}$ in
Eq.~(\ref{eq:f23}). To quadratic order in the momenta, the non-vanishing
string scattering amplitudes involving  only these two fields are:
\begin{eqnarray}
{\cal A}(\Phi_{23},{\bar\Phi}_{23},{\bar\Phi}_{23}^\dagger,
\Phi_{23}^\dagger)&=&
{\cal A}({\bar\Phi}_{23},\Phi_{23},\Phi_{23}^\dagger,
{\bar\Phi}_{23}^\dagger)=
{g^2\over 4}\left({su\over t}-16{s\over t}\right),\label{eq:xx1}\\
{\cal A}(\Phi_{23},{\bar\Phi}_{23},\Phi_{23}^\dagger,
{\bar\Phi}_{23}^\dagger)&=&
{\cal A}({\bar\Phi}_{23},\Phi_{23},{\bar\Phi}_{23}^\dagger,
\Phi_{23}^\dagger)=
{g^2\over 4}\left({st\over u}-16{s\over u}\right),\label{eq:xx2}\\
{\cal A}(\Phi_{23},\Phi_{23},\Phi_{23}^\dagger,\Phi_{23}^\dagger)&=&
{\cal A}({\bar\Phi}_{23},{\bar\Phi}_{23},{\bar\Phi}_{23}^\dagger,
{\bar\Phi}_{23}^\dagger)=
{g^2\over 4}\left({st\over u}+{su\over t}+2s
+16{s\over u}+16{s\over t}\right).\nonumber\\
\label{eq:xx3}
\end{eqnarray}
In Eqs.~(\ref{eq:xx1})--(\ref{eq:xx3}), the terms proportional to $s/t$
and $s/u$ are ``D-terms" due to the relevant gauge interactions ($\Phi_{23}$
and $\bar\Phi_{23}$ are charged under two $U(1)$'s) \cite{Smx2}. The presence
of such terms indicates that the scalar potential in the $\Phi_{23}$ and
$\bar\Phi_{23}$ directions is not flat, which is consistent with our
observation that these fields are not moduli. A simple comparison of
Eqs.~(\ref{eq:xx1})--(\ref{eq:xx3}) with Eq.~(\ref{eq:asugra})
does not yield something like Eq.~(\ref{eq:kd}), hence $\Phi_{23}$ and
${\bar\Phi}_{23}$ do not correspond to the canonical coordinates $\alpha_i$ of
the coset space $SO(2,1+12)/[SO(2)\times SO(1+12)]$; the ``telltale" sign is
the ``$2s$" in Eq.~(\ref{eq:xx3}). In this case, as we stated above, the
correct canonical coordinates associated with $\Phi_{23}$ and ${\bar\Phi}_{23}$
are ($\alpha_1=\Phi_1$)
\begin{equation}
\alpha_2={1\over\sqrt{2}}(\Phi_{23}+{\bar\Phi}_{23}),\quad
\alpha_3={-i\over\sqrt{2}}(\Phi_{23}-{\bar\Phi}_{23}).
\label{eq:ccc0}
\end{equation}
Indeed, in terms of $\alpha_2$ and $\alpha_3$ the string scattering amplitudes
(\ref{eq:xx1})--(\ref{eq:xx3}) become
\begin{eqnarray}
{\cal A}(\alpha_2,\alpha_3,\alpha_3^\dagger,\alpha_2^\dagger)&=&
{\cal A}(\alpha_3,\alpha_2,\alpha_2^\dagger,\alpha_3^\dagger)=
{g^2\over 4}\left({su\over t}+s+16{s\over u}\right),\label{eq:yy1}\\
{\cal A}(\alpha_2,\alpha_3,\alpha_2^\dagger,\alpha_3^\dagger)&=&
{\cal A}(\alpha_3,\alpha_2,\alpha_3^\dagger,\alpha_2^\dagger)=
{g^2\over 4}\left({st\over u}+s+16{s\over t}\right),\label{eq:yy2}\\
{\cal A}(\alpha_2,\alpha_2,\alpha_2^\dagger,\alpha_2^\dagger)&=&
{\cal A}(\alpha_3,\alpha_3,\alpha_3^\dagger,\alpha_3^\dagger)=
{g^2\over 4}\left({st\over u}+{su\over t}+s\right),\label{eq:yy3}\\
{\cal A}(\alpha_2,\alpha_2,\alpha_3^\dagger,\alpha_3^\dagger)&=&
{\cal A}(\alpha_3,\alpha_3,\alpha_2^\dagger,\alpha_2^\dagger)=
{g^2\over 4}\left(-s-16{s\over u}-16{s\over t}\right).\label{eq:yy4}
\end{eqnarray}
Note that the overall factor $(-i)$ in the definition of $\alpha_3$
in Eq.~(\ref{eq:ccc0}) is crucial in order to get the right sign in
front of the ``$s$" term in Eq.~(\ref{eq:yy4}) ({\em c.f.}
Eq.~(\ref{eq:ap4p})). In addition to (\ref{eq:yy1})--(\ref{eq:yy4}), we also
have the  following non-vanishing string scattering amplitudes involving the
modulus field $\alpha_1$ and the matter fields $\alpha_i\ (i=2,3)$:
\begin{eqnarray}
{\cal A}(\alpha_1,\alpha_i,\alpha_i^\dagger,\alpha_1^\dagger)&=&
{\cal A}(\alpha_i,\alpha_1,\alpha_1^\dagger,\alpha_i^\dagger)=
{g^2\over 4}\left({su\over t}+s\right),\label{eq:zz1}\\
{\cal A}(\alpha_1,\alpha_i,\alpha_1^\dagger,\alpha_i^\dagger)&=&
{\cal A}(\alpha_i,\alpha_1,\alpha_i^\dagger,\alpha_1^\dagger)=
{g^2\over 4}\left({st\over u}+s\right),\label{eq:zz2}\\
{\cal A}(\alpha_1,\alpha_1,\alpha_i^\dagger,\alpha_i^\dagger)&=&
{\cal A}(\alpha_i,\alpha_i,\alpha_1^\dagger,\alpha_1^\dagger)=
{g^2\over 4}(-s).\label{eq:zz3}
\end{eqnarray}
The comparison between Eqs.~(\ref{eq:yy1})--(\ref{eq:zz3}) with
Eq.~(\ref{eq:asugra}) yields Eq.~(\ref{eq:kd}), which demonstrates
that $\alpha_{2,3}$ defined in (\ref{eq:ccc0}) are indeed the correct
canonical coordinates.

The above analysis can be readily extended to the matter fields $h_1$ and
${\bar h}_1$, with similar results, \ie, the canonical coordinates are
${1\over\sqrt{2}}(h_1+\bar h_1)$ and ${-i\over\sqrt{2}}(h_1-\bar h_1)$
(five components each). To write down the K\"ahler potential for the coset
space $SO(2,1+12)/[SO(2)\times SO(1+12)]$ spanned by the scalar fields
$\Phi_1$, $\Phi_{23}$, ${\bar\Phi}_{23},h_1,{\bar h}_1$, we begin with the
canonical form (\ref{eq:kkk}) using the properly defined complex coordinates
$\alpha_i$, and then rewrite all the $\alpha_i$ in terms of the original string
states. In our current example we obtain
\begin{eqnarray}
K^{(1)}(\Phi,\Phi^\dagger)&=&-\log\left(1-\Phi_1\Phi_1^\dagger
-\Phi_{23}\Phi_{23}^\dagger-\bar\Phi_{23}\bar\Phi_{23}^\dagger
-h_1h_1^\dagger-{\bar h}_1{\bar h}_1^\dagger\right. \nonumber\\
&&\left.+{1\over 4}|\Phi_1^2+2\Phi_{23}\bar\Phi_{23}+2h_1{\bar h}_1|^2\right),
\label{eq:k1f5}
\end{eqnarray}
where we have suppressed the $SU(5)$ group indices for $h_1$ and
${\bar h}_1$. Thus we see that those string scattering amplitudes
in terms of the original string states, \ie,
Eqs.~(\ref{eq:xx1})--(\ref{eq:xx3}) are indeed consistent
with (\ref{eq:k1f5}). The canonical coordinates $\alpha_i$
introduced above help provide a systematic proof of this result,
and can be discarded once this is accomplished. In
practice,  by simply following the above example, one can easily work out
the K\"ahler potential for all the untwisted scalar matter fields.

In the ``revamped'' flipped $SU(5)$ model, the {\em second set} of scalar
fields from the Neveu-Schwarz sector [$\Phi_2,\Phi_{31},\bar\Phi_{31},h_2,\bar
h_2$] parameterize a coset space $SO(2,1+12)/[SO(2)\times SO(1+12)]$,
whose K\"ahler potential is given by
\begin{eqnarray}
K^{(2)}(\Phi,\Phi^\dagger)&=&-\log\left(1-\Phi_2\Phi_2^\dagger
-\Phi_{31}\Phi_{31}^\dagger-\bar\Phi_{31}\bar\Phi_{31}^\dagger
-h_2h_2^\dagger-{\bar h}_2{\bar h}_2^\dagger\right. \nonumber\\
&&\left.+{1\over 4}|\Phi_2^2+2\Phi_{31}\bar\Phi_{31}+2h_2{\bar h}_2|^2\right),
\label{eq:k1f5x}
\end{eqnarray}
where $\Phi_2=H^{(2)}_2$ is the modulus field (see Sec.~\ref{realistic}).
The {\em third set} [$\Phi_3,\Phi_4,\Phi_5,\Phi_{12},\bar\Phi_{12}$, $h_3,\bar
h_3$] contains two moduli fields $\Phi_{4,5}=H^{(3)}_{1,2}$ (see
Sec.~\ref{realistic}) and thirteen matter fields, which parameterize a coset
space $SO(2,2+13)/[SO(2)\times SO(2+13)]$ with the following K\"ahler
potential
\begin{eqnarray}
K^{(3)}(\Phi,\Phi^\dagger)&=&-\log\left(1-\Phi_4\Phi_4^\dagger
-\Phi_5\Phi_5^\dagger-\Phi_3\Phi_3^\dagger
-\Phi_{12}\Phi_{12}^\dagger-\bar\Phi_{12}\bar\Phi_{12}^\dagger
-h_3h_3^\dagger-{\bar h}_3{\bar h}_3^\dagger\right. \nonumber\\
&&\left.+{1\over 4}|\Phi_4^2+\Phi_5^2+\Phi_3^2
+2\Phi_{12}\bar\Phi_{12}+2h_3{\bar h}_3|^2\right),
\label{eq:k1f5y}
\end{eqnarray}
where $\Phi_3$ is a matter field whose right-moving fermionic oscillator is
{\em real} and transforms as a singlet under $SU(5)\times U(1)$ and has
no $U(1)$ charges (see Ref.~\cite{fsu5re}).

Finally we note that we can confirm in string pertubation theory that the
potential is flat in the moduli directions. In this case, since the moduli are
neutral under all gauge symmetries they do not appear in $D$-terms. Moreover,
restricting the superpotential of the model to terms involving only products of
three untwisted fields, one can verify that the moduli do not appear as
$F$-terms either: the only such terms are
$(\Phi_{12}\Phi_{23}\Phi_{31}+\bar\Phi_{12}\bar\Phi_{23}\bar\Phi_{31})$
\cite{fsu5re}.

Although here we only discuss
explicitly the ``revamped'' flipped $SU(5)$ model, it is a straightforward
exercise to apply our method to any fermionic string model of this class, such
as those derived in Refs.~\cite{pati,smlike,Search}.

\section{Target space duality invariance}
\label{target}
When the moduli fields move around in the moduli space ${\cal M}$, their
associated exactly marginal operators generate deformations of the underlying
CFT of the string model. One well-known stringy phenomenon is that a subset of
these deformations leads to new CFTs which are physically equivalent to the
original one. Such deformations correspond to some discrete reparameterizations
of the moduli space ${\cal M}$, which are referred to as the target space
duality transformations, and form the discrete duality group $\Gamma$ under
which the string spectrum is invariant \cite{ta1,ta2,ta3,ta4}. The target space
duality invariance strongly restricts the K\"ahler potential and the
superpotential of the string-derived effective field theory \cite{ta4}. In this
section, based on the results obtained in previous sections, we discuss the
target space duality invariance in the context of fermionic string models, in
particular, we establish the properties of the untwisted matter fields under
the target space duality transformations.

As we have shown, for the class of fermionic string models that we
are considering, the total untwisted moduli space factorizes
into three distinct subspaces because the moduli fields separate into
three sets. The sub-moduli space is either $SO(2,2)/SO(2)\times SO(2)$
when there are two moduli fields, or $SO(2,1)/SO(2)$ when there is
only one modulus field (see Eqs.~(\ref{eq:space2}) and (\ref{eq:space3})).
In what follows, we examine these two cases separately.

We first consider the case of moduli space $SO(2,2)/SO(2)\times SO(2)$.
According to the discussion in Sec.~\ref{z2}, one can replace the two
moduli fields $H_{1,2}$ by a $T$-type moduli $\Phi_T$ and a $U$-type
moduli $\Phi_U$, such that each separately parameterizes a coset space
$SU(1,1)/U(1)$. The target space duality group in this case is given by
$PSL(2,{\bf Z})_T\times PSL(2,{\bf Z})_U$ \cite{ta2,IBLust}, acting on the
moduli fields $T,U$ in the so-called ``supergravity basis'' as
\begin{equation}
T\rightarrow {{a_TT-ib_T}\over {ic_TT+d_T}},\quad
U\rightarrow {{a_UU-ib_U}\over {ic_UU+d_U}},
\label{eq:modutr1}
\end{equation}
where $a,b,c,d\in {\bf Z}$ and $ad-bc=1$ for the coefficients of the
$T$-transformation and the $U$-transformation. The moduli fields $\Phi_{T,U}$
in the ``string basis'' are related to the moduli fields $T,U$ in the
``supergravity basis" through the following equations \cite{ILLT}
\begin{equation}
\Phi_T={{T_c-T}\over {{\bar T}_c+T}},\quad
\Phi_U={{U_c-U}\over {{\bar U}_c+U}},
\label{eq:defxx}
\end{equation}
where $T_c(U_c)$ is an unspecified complex number which can be viewed as the
vacuum expectation value of field $T$ in the {\it free} fermionic model that we
start with, which corresponds to a critical point in the moduli space.
{}From Eq.~(\ref{eq:defxx}) we see that at this critical point
$\langle\Phi_T\rangle=\langle\Phi_U\rangle=0$, which is consistent with the
fact that all Thirring interactions are turned
off in the {\it free} fermionic model.

Now suppose that in addition to the moduli fields $\Phi_T$ and $\Phi_U$ there
are $n$ associated untwisted matter fields, which for convenience we express in
terms of the corresponding canonical coordinates $\alpha_i$. Starting from
Eq.~(\ref{eq:kkk}) for the case with $n+2$ coordinates, but replacing two of
them (the moduli $H_{1,2}$) with fields $\Phi_{T,U}$ according to
Eqs.~(\ref{eq:T}) and (\ref{eq:U}), we can write the full K\"ahler
potential for all these untwisted fields as
\begin{eqnarray}
K&=&-\log\left\{(1-\Phi_T{\bar\Phi}_T)(1-\Phi_U{\bar\Phi}_U)
-\sum^n_i\alpha_i{\bar\alpha}_i
+{1\over 4}\left|\sum^n_i\alpha^2_i\right|^2\right. \nonumber\\
&&\qquad\qquad\left.+{1\over 2}\Phi_T\Phi_U\sum^n_i{\bar\alpha}^2_i
+{1\over 2}{\bar\Phi}_T{\bar\Phi}_U\sum^n_i\alpha^2_i\right\}.
\label{eq:wkp1}
\end{eqnarray}
We now show that the K\"ahler potential (\ref{eq:wkp1}) is target space duality
invariant. First we recall that the physical content of target space duality
{\it not only} includes the transformations (\ref{eq:modutr1}) for the moduli
fields $T$ and $U$, which are the generalizations of the famous
$R\rightarrow 1/2R$ duality transformation,
{\it but also} requires the simultaneous interchange of
the Kaluza-Klein (momentum) modes with the winding modes. This interchange is
equivalent \cite{ILLT} to transforming the critical values $T_c$ and $U_c$
in the same way as the fields $T$ and $U$ according to (\ref{eq:modutr1}).
Therefore, from (\ref{eq:defxx}) we see that the moduli fields $\Phi_{T,U}$ in
the ``string basis'' simply transform under target space duality group
$PSL(2,{\bf Z})_T\times PSL(2,{\bf Z})_U$ by a field-independent phase, namely,
\begin{equation}
\Phi_T\rightarrow e^{-2i\arg(ic_TT_c+d_T)}\,\Phi_T,\quad
\Phi_U\rightarrow e^{-2i\arg(ic_UU_c+d_U)}\,\Phi_U.
\label{eq:modutr2}
\end{equation}
Given the transformations (\ref{eq:modutr2}) for the moduli $\Phi_{T,U}$, in
order for the K\"ahler potential (\ref{eq:wkp1}) to be target space duality
invariant, it is not hard to see that the matter fields $\alpha_i$ have to
transform universally as
\begin{equation}
\alpha_i\rightarrow e^{-i\arg(ic_TT_c+d_T)}\,
e^{-i\arg(ic_UU_c+d_U)}\,\alpha_i\ .
\label{eq:mmtr}
\end{equation}
Since the canonical coordinates $\alpha_i$ are related to the string matter
fields through linear transformations (see \eg, Eq.~(\ref{eq:ccc0})),
this argument shows that the string matter fields transform under target space
duality also according to (\ref{eq:mmtr}).

Two remarks are in order. First, we see that the matter fields
{\it do not} transform as modular forms. This is because we are working in
the ``string basis'', whereas concepts such as modular forms only come
into play if the duality properties are analyzed in the so-called
``supergravity basis", as it has been done traditionally \cite{ta4}.
Second, we have demonstrated the target space duality invariance
of (\ref{eq:wkp1}) {\it without} performing any additional K\"ahler
transformations. Again, this is a result of working in the
``string basis'', where the fields are basically ``inert" under modular
transformations (up to field-independent phases). In the ``string basis'',
since the K\"ahler potential $K$ is itself target space duality invariant, the
requirement of an invariant K\"ahler {\em function} $G=K+\log|W|^2$ implies
that the superpotential $W$ (written in terms of the ``string basis" fields)
can only be allowed to have a phase transformation, \ie, $W\to
e^{i\varphi}\,W$. In fact, one can show \cite{KLN} that in non-vanishing cubic
terms in $W$ which are products of three untwisted fields, each field comes
from a different set, \ie,
\begin{equation}
W=\sum \lambda_{ijk}\, \alpha^{(1)}_i\alpha^{(2)}_j\alpha^{(3)}_k\ .
\label{W}
\end{equation}
Thus, suppose the sub-moduli space of the first set is
$SO(2,2)/SO(2)\times SO(2)$, then from Eq.~(\ref{eq:mmtr}) we see that
under target space duality group
$PSL(2,{\bf Z})_T\times PSL(2,{\bf Z})_U$ associated with this set,
\begin{equation}
W\to\sum\lambda_{ijk}\, e^{-i\varphi^{(1)}_T}\,e^{-i\varphi^{(1)}_U}
\alpha^{(1)}_i\alpha^{(2)}_j\alpha^{(3)}_k =
e^{-i\varphi^{(1)}_T-i\varphi^{(1)}_U}W,
\label{Wtransf}
\end{equation}
as required by duality invariance of $G$. Here
$\varphi^{(1)}_T=\arg(ic_T T_c+d_T)$ and $\varphi^{(1)}_U=\arg(ic_U U_c+d_U)$.
This assumption also shows that moduli (which tranform as in
Eq.~(\ref{eq:modutr2})) are not allowed in all-untwisted-field cubic couplings,
as expected from their flatness properties. Extension of this argument to
non-renormalizable terms leads to non-trivial constraints on the corresponding
couplings, which acquire a non-trivial moduli dependence \cite{modinv}.

To make contact with previous results, we also study this problem in the
``supergravity basis". To this end, one starts with the moduli fields
$T,U$ instead of $\Phi_{T,U}$, and then it is {\it necessary} to
make a holomorphic field redefinition for the matter fields such that
the ``supergravity basis'' matter fields ($A_i$) are given by their
``string basis" counterparts ($\alpha_i$) as follows
\begin{equation}
A_i={{{\bar T}_c+T}\over\sqrt{T_c+{\bar T}_c}}
{{{\bar U}_c+U}\over\sqrt{U_c+{\bar U}_c}}\,\alpha_i.
\label{eq:str_sugra}
\end{equation}
In this ``supergravity basis'', up to a K\"ahler transformation,
one can bring (\ref{eq:wkp1}) into the following form
\begin{eqnarray}
K'&=&-\log\left\{(T+{\bar T})(U+{\bar U})
-\sum^n_iA_i{\bar A}_i
+{1\over 4}{(T_c+{\bar T}_c)\over |{\bar T}_c+T|^2}
{(U_c+{\bar U}_c)\over |{\bar U}_c+U|^2}\left|\sum^n_iA^2_i\right|^2
\right. \nonumber\\
&&\left.+{1\over 2}\left({{T_c-T}\over {T_c+{\bar T}}}\right)
\left({{U_c-U}\over {U_c+{\bar U}}}\right)\sum^n_i{\bar A}^2_i
+{1\over 2}\left({{{\bar T}_c-{\bar T}}\over {{\bar T}_c+T}}\right)
\left({{{\bar U}_c-{\bar U}}\over {{\bar U}_c+U}}\right)\sum^n_iA^2_i\right\},
\label{eq:wkp2}
\end{eqnarray}
which under target space duality transformations (\ref{eq:modutr1}) for the
moduli fields $T,U$ and the critical values $T_c,U_c$ transforms to
\begin{equation}
K'\rightarrow K'+\log(|ic_TT+d_T|^2|ic_UU+d_U|^2),
\label{eq:www}
\end{equation}
accompanied by the
the following universal transformation for matter fields $A_i$,
\begin{equation}
A_i\rightarrow {1\over (ic_TT+d_T)(ic_UU+d_U)}\,A_i.
\label{eq:weight}
\end{equation}
Note that the last two terms in (\ref{eq:wkp2}) fix completely the phase in
(\ref{eq:weight}), and thus the $A_i$ transform as a modular form of weight
($-1$). The duality transformation (\ref{eq:weight}) can also be readily
obtained from Eq.(\ref{eq:str_sugra}).
In this ``supergravity basis'' the superpotential $W$ must also
transform as a modular form of weight ($-1$) in order to cancel the
second term  in (\ref{eq:www}). This is the usual analysis of
Refs.~\cite{ta4,IBLust}. However, in our case, at least for the discussion of
target space duality, such an approach does not seem to be necessary.
In fact, the presence of the second term in (\ref{eq:www}) is because
one has neglected a K\"ahler transformation when bringing $K$
of (\ref{eq:wkp1}) into $K'$ of (\ref{eq:wkp2}); this piece
automatically cancels the second term in (\ref{eq:www}) so that
$K$ is completely invariant, as we have shown above in terms of
the ``string basis''.

We next discuss the case of moduli space $SO(2,1)/SO(2)$. In this case,
according to (\ref{eq:nonli}), we can replace the moduli field $H$ by
$\Phi_t$ which leads to the Fubini-Study metric of $SU(1,1)/U(1)$.
The target space duality group is simply $PSL(2,{\bf Z})$, under which
the modulus field $t$, in the ``supergravity basis'', transforms as
\begin{equation}
t\rightarrow {{a_tt-ib_t}\over {ic_tt+d_t}},
\label{eq:modutr3}
\end{equation}
with $a_t,b_t,c_t,d_t\in {\bf Z}$ and $a_td_t-b_tc_t=1$.
The field $\Phi_t$, although not a string state in this case,
can still be related to $t$ via
\begin{equation}
\Phi_t={{t_c-t}\over {{\bar t}_c+t}},
\label{eq:defxx0}
\end{equation}
where $t_c$ is the critical value of $t$ in the {\it free} fermionic
model. Therefore, under the target space duality group $PSL(2,{\bf Z})$,
$\Phi_t$  simply goes through a field-independent phase transformation,
\begin{equation}
\Phi_t\rightarrow e^{-2i\arg(ic_tt_c+d_t)}\,\Phi_t.
\label{eq:modutr4}
\end{equation}
This is also the transformation for the original string state $H$,
as can be seen from (\ref{eq:nonli}). In this sense, we can roughly
refer to $\Phi_t$ as the modulus field in the ``string basis''.

In this case, starting from Eq.~(\ref{eq:kkk}) for the case with
$n+1$ coordinates, but replacing one of them (the modulus $H$) by
$\Phi_t$ using the inverse of Eq.~(\ref{eq:nonli}), one can obtain the
following full K\"ahler potential
\begin{eqnarray}
K&=&-\log\left\{1-\Phi_t{\bar\Phi}_t-\sum^n_i\alpha_i{\bar\alpha}_i
+{1\over 4}\left|\sum^n_i\alpha^2_i\right|^2\right. \nonumber\\
&&\left.+{1\over 2}\,{1\over 1+\sqrt{1-\Phi_t{\bar\Phi}_t}}
\left(\Phi_t^2\sum^n_i{\bar\alpha}^2_i
+{\bar\Phi}_t^2\sum^n_i\alpha^2_i\right)\right\},
\label{eq:ykp1}
\end{eqnarray}
which is target space duality invariant provided the matter
fields transform universally as
\begin{equation}
\alpha_i\rightarrow e^{-2i\arg(ic_tt_c+d_t)}\,\alpha_i.
\label{eq:mmtr2}
\end{equation}
We note that in this case the matter fields $\alpha_i$ transform in the same
way as the modulus field $\Phi_t$ (see Eq.~(\ref{eq:modutr4})), whereas in the
case of moduli space $SO(2,2)/SO(2)\times SO(2)$ this is not true (see
Eqs.~(\ref{eq:modutr2}) and (\ref{eq:mmtr})).
One can also carry out the usual analysis of Refs.~\cite{ta4,IBLust}
for the case of moduli space $SO(2,1)/SO(2)$, as we did for the case
of moduli space $SO(2,2)/SO(2)\times SO(2)$. The only difference
is that the holomorphic field redefinition which defines the matter
fields in the ``supergravity basis'' here becomes
\begin{equation}
A_i={{{\bar t}_c+t}\over\sqrt{t_c+{\bar t}_c}}\,\alpha_i,
\end{equation}
and the $A_i$ transform as modular forms of weight ($-1$) with a
non-trivial phase
\begin{equation}
A_i\rightarrow {e^{-i\arg(ic_tt_c+d_t)}\over (ic_tt+d_t)}\, A_i,
\end{equation}
which should be contrasted with (\ref{eq:weight}) for the
case of moduli space $SO(2,2)/SO(2)\times SO(2)$.

\section{Conclusions}
\label{conclusions}
Identifying the moduli fields and their symmetries in string-derived models
constitutes the first step in the determination of the low-energy effective
field theory. This knowledge allows one to calculate the K\"ahler potential,
which together with the superpotential and gauge kinetic functions determine
completely the effective supergravity theory. This effective theory can then
be used for various phenomenological studies, such as string threshold
corrections in gauge coupling unification, supersymmetry breaking, string
cosmology, etc.

We have presented a general procedure by which the untwisted moduli fields
in fermionic models can be identified. The crucial element of our procedure
is the deformation of the free-fermionic model by the exactly marginal
operators which take the form of world-sheet Abelian Thirring interactions,
thus getting away from the free-fermionic point to appreciate the duality
symmetries embodied in the moduli fields. We also back all of our generic
symmetry-based arguments by explicit string perturbation theory calculations.

Previous phenomenological studies in free-fermionic models
\cite{thresholds,condensates,modinv,Alonnew} assumed that the usual
symmetric orbifold analysis was applicable. Our results show that this
assumption holds only for the simplest free-fermionic models,
but fails in the case of realistic models where the asymmetric nature
of the equivalent orbifold formulation is essential. Futhermore,
with several examples we have demonstrated the method to obtain
the full K\"ahler potential for the non-linear sigma model of the
untwisted sector in such fermionic models, in terms of the {\em actual
massless string states}; this problem had remained largely obscure until
now.

The results derived in this paper concerning the moduli space and
the K\"ahler function, \eg, Eqs.~(\ref{eq:stspc}), (\ref{eq:kp1}) and
(\ref{eq:stf5}), (\ref{eq:kp2}), bear close resemblance to results
obtained early on in the context of {\em no-scale} supergravity
\cite{EKNI+II,LN}. In particular, the ever-present moduli fields in
string models lead to flat potentials which are characteristic of
no-scale supergravity. The corresponding expressions for the K\"ahler
function in our string-derived supergravity are more complex than those
obtained in traditional no-scale supergravity, because of the rich
string-theory structure underlying the effective supergravity model.
Nonetheless, the original motivations embodied in no-scale
supergravity, \ie, the vanishing of the cosmological constant and the
flat potentials which allow the dynamical determination of mass scales
\cite{Lahanas}, remain valid and we would hope that such a physics program
could also be pursued successfully in the context of {\em string} no-scale
supergravity.

\section*{Acknowledgments}

This work has been supported in part by DOE grant DE-FG05-91-ER-40633.
The work of K.Y. has been supported by the World Laboratory. K.Y.
would like to thank Ioannis Giannakis for stimulating discussions,
and Lance Dixon for an e-mail correspondence.

\vfill\eject

\vfill\eject
\end{document}